\documentclass[a4paper,11pt]{article}
\usepackage{jcappub}
\usepackage{hyperref}
\usepackage{bm}
\usepackage{float}
\usepackage[nooneline]{subfigure}
\restylefloat{figure}

\title{\boldmath Constraining the Minimum Luminosity of High Redshift Galaxies through Gravitational Lensing}

\author{Natalie Mashian}
\author{and Abraham Loeb}
\affiliation{Harvard-Smithsonian Center for Astrophysics, \\60 Garden Street, Cambridge, MA 02138, USA}
\emailAdd{nmashian@physics.harvard.edu}
\emailAdd{aloeb@cfa.harvard.edu}

\abstract{We simulate the effects of gravitational lensing on the source count of high redshift galaxies as projected to be observed by the Hubble Frontier Fields program and the James Webb Space Telescope (JWST) in the near future. Taking the mass density profile of the lensing object to be the singular isothermal sphere (SIS) or the Navarro-Frenk-White (NFW) profile, we model a lens residing at a redshift of $z_L$ = 0.5 and explore the radial dependence of the resulting magnification bias and its variability with the velocity dispersion of the lens, the photometric sensitivity of the instrument, the redshift of the background source population, and the intrinsic maximum absolute magnitude ($M_{max}$) of the sources.  We find that gravitational lensing enhances the number of galaxies with redshifts $z \gtrsim$ 13 detected in the angular region $\theta_E/2 \leq \theta \leq 2\theta_E$ (where $\theta_E$ is the Einstein angle) by a factor of $\sim$ 3 and 1.5 in the HUDF ($df/d\nu_0 \sim$ 9 nJy) and medium-deep JWST surveys ($df/d\nu_0 \sim$ 6 nJy). Furthermore, we find that even in cases where a negative magnification bias reduces the observed number count of background sources, the lensing effect improves the sensitivity of the count to the intrinsic faint-magnitude cut-off of the Schechter luminosity function. In a field centered on a strong lensing cluster, observations of $z \gtrsim$ 6 and $z \gtrsim$ 13 galaxies with JWST can be used to infer this cut-off magnitude for values as faint as $M_{max} \sim$ -14.4 and -16.1 mag ($L_{min} \approx$ 2.5$\times$10$^{26}$ and 1.2$\times$10$^{27}$ erg s$^{-1}$ Hz$^{-1}$) respectively, within the range bracketed by existing theoretical models. Gravitational lensing may therefore offer an effective way of constraining the low-luminosity cut-off of high-redshift galaxies.}


\begin{document}
\maketitle

\section{Introduction}
\label{1}
The characterization of the earliest galaxies in the universe remains one of the most important frontiers of observational cosmology, and also one of the most challenging~\cite{1}. High-redshift searches carried out with the \emph{Hubble Space Telescope} (HST) have recently provided significant insights to the mass assembly and buildup of the earliest galaxies ($z \gtrsim$ 6) and the contribution of star formation to cosmic reionization~\cite{2,3,4,5}. However, because of their great distances and extreme faintness, as well as the high sky background, high redshift galaxies remain difficult to detect. Furthermore, those sources which are  bright enough to be studied individually are drawn from the bright tail of the luminosity function (LF) of high redshift galaxies and are therefore not necessarily representative of the bulk of the population~\cite{6}. Gravitational lensing by galaxy clusters has been highlighted as an efficient way of improving this situation, providing an opportunity to observe the high-redshift universe in unprecedented detail~\cite{7,8}. 
\\ \indent Light rays propagating through the inhomogeneous gravitational field of the Universe are often deflected by intervening clumps of matter, which cause most sources to appear slightly displaced and distorted in comparison with the way they would otherwise appear in a perfectly homogeneous and isotropic universe~\cite{9,10,11}. When the light from a distant galaxy is deflected by foreground mass concentrations such as galaxies, groups, and galaxy clusters, its angular size and brightness are increased and multiple images of the same source may form. This phenomenon, referred to as \emph{strong gravitational lensing}, leads to a magnification bias that can have a significant effect on the observability of a population of galaxies. Magnified sources, that would otherwise be too faint for detection without a huge investment of observing time, can be found, and unresolved substructure and morphological details in these intrinsically faint galaxies can be studied~\citep{12,13,14}. The light magnification produced by nature's "cosmic telescopes" can be exploited in the study of high-redshift galaxies which have greater probability of falling in alignment with, and therefore being lensed by, a foreground galaxy~\citep{15}.~\cite{16} explored the prospect of detecting a hypothetical population of population III galaxies via gravitational lensing by a particular galaxy cluster (MACS J0717.5 + 3745) as the lens. Indeed, several highly magnified galaxy candidates at up to redshift $z \sim$ 10 have already been discovered behind massive clusters~\cite{6,7,8,17,18,19,20,21}. The Hubble Frontier Fields program\footnote{http://www.stsci.edu/hst/campaigns/frontier-fields/} is expected to lead to many more such discoveries. With its six deep fields centered on strong lensing galaxy clusters in parallel with six deep ``blank fields", the Hubble Frontier Fields will reveal previously inaccessible populations of $z$ = 5-10 galaxies that are 10-50 times intrinsically fainter than any presently known.  In the coming decade, the planned \emph{James Webb Space Telescope} (JWST)\footnote{http://www.stsci.edu/jwst/} promises to go even further by placing new constraints on the stellar initial mass function at high redshift, on the luminosity function of the first galaxies, and on the progress of the early stages of reionization with observations of galaxies at $z \gtrsim$ 10~\cite{22,23,24}. 
\\ \indent While lenses magnify the observed flux and lift sources which are intrinsically too faint to be observed over the detection threshold, they simultaneously increase the solid angle within which sources are observed and thus reduce their number density and measured surface brightness in the sky~\cite{25}.  Zemcov et al. recently reported measuring a deficit of surface brightness within the central region of several massive galaxy clusters with the SPIRE instrument, and used the deficit to constrain the surface brightness of the cosmic infrared background~\cite{26}. The outcome of this trade-off between depth and area depends on a variety of factors, such as the photometric sensitivity of the detecting device and the slope of the luminosity function of background sources. Given a photometric sensitivity capable of detecting faint sources even in the absence of any light amplification, the lensing effect leads to a negative magnification bias, reducing the apparent surface density behind lensing clusters. If, however, the fainter sources cannot be observed unless magnified, then whether the magnification bias leads to a surplus or deficit of observed sources depends on the effective slope of their luminosity function~\cite{27}. At fainter magnitudes where the effective slope, $\alpha$, is shallow, there may not be enough faint sources in the lensed population to compensate for the increase in total surface area. However, in cases where $\alpha \gtrsim$ 2, the gain in depth due to apparent brightening may outweigh the loss in area;  gravitational lensing will thus increase the apparent surface density behind the lensing object, boosting up the number of detected sources relative to that which would otherwise be observed in an unlensed field~\cite{20,28}.
\\\ \indent The observed number counts of galaxies residing at redshifts greater than some $z'$ may also be sensitive to the intrinsic faint-magnitude cut-off chosen for the extrapolation of the galaxy LF. Theoretical and numerical investigations have established that a halo at $z \lesssim$ 10 irradiated by a UV field comparable to the one required for reionization needs a mass $M_h$ $\gtrsim$ (0.6 - 1.7) $\times$ 10$^8$ M$_\odot$, with a corresponding temperature $T_{vir}$ $\gtrsim$ (1 - 2)$\times$10$^{4}$ K at $z$ = 7, in order to cool and form stars~\cite{29,30,31}. Such claims have motivated models with cut-offs for the absolute magnitude of the smallest halo capable of forming stars as faint as $M_{AB} \approx$ -10~\cite{32,33}. Gravitational lensing may provide an effective way to constrain the value of this minimum luminosity given the fact that the lensed number count of high-redshift galaxies remains sensitive to this intrinsic low-luminosity cut-off at much fainter values compared to the observed number count in a blank field.
\\ \indent In this paper, we predict the lensing rate of high-redshift objects that will be observed with both HST Frontier Fields in the upcoming months, and JWST within the next decade. In section \ref{2} we consider two different axially symmetric lens models: a singular isothermal sphere and a NFW profile~\cite{34} lens for comparison, examining their respective effects on the number count of the background lensed galaxy population. In addition to considering lensing clusters, we also consider galaxy-group lensing and compute the lensing rates expected in each case given the velocity dispersion of the lensing object.  We present our numerical results in section \ref{3} and show the transition from a positive to a negative magnification bias as a function of the minimum intrinsic luminosity, the photometric sensitivity, and the angular distance from the given lens. We conclude in section \ref{4} with a discussion of our findings and their implications for observations with the HST Frontier Fields and JWST in the near future.Throughout this paper, we adopt $\Omega_m$ = 0.3 and $\Omega_\Lambda$ = 0.7 as the present-day density parameters of matter and vacuum, respectively and take $H_0$ = 100$h$ km s$^{-1}$ Mpc$^{-1}$ as the Hubble constant with $h$ = 0.7. We express all magnitudes in the AB system.

\section{The Lensing Model}
\label{2}
The ray-tracing equation that relates the position of a source, $\vec{\eta}$, to the impact parameter of a light ray in the lens plane, $\vec{\xi}$, is given in angular coordinates by
\begin{equation}
\vec{\beta} = \vec{\theta} - \vec{\alpha}(\vec{\theta})
\end{equation}
where $\vec{\beta}$ = $\vec{\eta}$/D$_{s}$, $\vec{\theta}$ = $\vec{\xi}$/D$_{l}$, and $\vec{\alpha}$ is the reduced deflection angle due to a lens with surface mass density $\Sigma$,
\begin{equation}
\vec{{\alpha}}(\vec{\theta}) = \frac{4G}{c^2}\frac{D_{ls}D_{l}}{D_{s}} \int \frac{(\vec{\theta}-\vec{\theta}^{\prime})\Sigma(\vec{\theta}^{\prime})}{|\vec{\theta}-\vec{\theta}^{\prime}|^2}d^2\theta^{\prime} \; .
\end{equation}
 $D_{l,s,ls}$ are the angular-diameter distances between observer and lens, observer and source, and lens and source, respectively. In the standard $\Lambda$CDM cosmology, the angular-diameter distance $D_{A}(z)$ of a source at redshift $z$ is defined as
\begin{equation}
D_{A}(z) = \frac{c}{H_0}\frac{1}{(1+z)}\int ^{z}_{0}\frac{dz'}{E(z')}
\end{equation}
where 
\begin{equation}
E(z) = \sqrt{\Omega_m(1+z)^3+\Omega_{\Lambda}} \;.
\end{equation}

For a circularly-symmetric mass distribution, $\Sigma(\vec{\theta}) = \Sigma(|\vec{\theta}|)$; the dimensionless surface mass density, also referred to as the convergence, is then given by

\begin{equation}
\kappa(x) = \frac{\Sigma(x)}{\Sigma_{cr}}\;\; \text{    with    } \;\;\Sigma_{cr}=\frac{c^2D_s}{4\pi GD_lD_{ls}}
\end{equation}
where we have introduced the dimensionless impact parameter $\vec{x}$ = $\vec{\theta}/\theta_0$ with an arbitrary angular scale $\theta_0$. The corresponding dimensionless mass $m(x)$ within a circle of angular radius $x$ is then
\begin{equation}
m(x) = 2\int_0^x dx'x'\kappa(x') 
\end{equation}
and the magnification factor in terms of these dimensionless quantities takes the following form,
\begin{equation}\label{2.7}
\mu(x) = \frac{1}{(1-\frac{m(x)}{x^2})(1+\frac{m(x)}{x^2}-2\kappa(x))} \;\;.
\end{equation} 

A simple model for the matter distribution in a gravitational lens is a Singular Isothermal Sphere (SIS)~\cite{9} with a surface mass density of
\begin{equation}
\Sigma_{SIS}(\xi) = \frac{\sigma_v^2}{2G\xi} \;\;.
\end{equation}
 where $\sigma_v$ is the line-of-sight velocity dispersion of the lens. In this case, the reduced deflection angle, commonly referred to as the Einstein angle for the SIS lens and denoted as $\theta_E$, is independent of the impact parameter,
 \begin{equation}
 \theta_E = 4\pi\frac{\sigma_v^2}{c^2}\frac{D_{ls}}{D_s}
 \end{equation}
 and the lens equation reduces to 
 \begin{equation}
\vec{\beta} = \vec{\theta} - \vec{\alpha}(\vec{\theta}) =  \vec{\theta} -\frac{ \vec{\theta}}{|\vec{\theta}|}\theta_E
\end{equation}
where negative angles refer to positions on the opposite side of the lens center. The lensing effect causes the image of the source to be displaced, magnified, and sometimes split~\cite{25}. When $|\beta|$ $<$ $\theta_E$, the lens equation has two solutions, $\theta_\pm$ = $\beta$ $\pm$ $\theta_E$, and multiple images are obtained. Conversely, if the source lies outside the Einstein ring, i.e. $|\beta|$ $>$ $\theta_E$, only one image is present at $\theta$ = $\theta_+$ = $\beta$ + $\theta_E$.
The corresponding magnification factor due to a SIS lens is given by
\begin{equation}
\mu_{SIS}(\theta) = \left(1-\frac{\theta_E}{|\theta|}\right)^{-1}
\end{equation}
where negative values of $\mu$ correspond to inverted images. For large values of $\theta$, $\mu \approx$ 1 and the source is weakly affected by the lensing potential, while for $\theta$ = $\theta_E$, the magnification diverges, corresponding to the formation of an Einstein ring. In practice, the maximum magnification is limited by the finite extent of the lensed source~\cite{35}. Since we are considering primarily the lensing effect on compact, high-redshift galaxies~\cite{36}, we ignore the angular size of the sources and model the background as a collection of point sources. 

Although the SIS model is useful in providing a good first-order approximation to the projected mass distribution of known early-type galaxies and cluster lenses~\cite{37,38,39,40,41}, it is not an entirely realistic model. In particular, Meneghetti et al. finds that the contributions of ellipticity, asymmetries, and substructures amount to $\sim$40\%, $\sim$10\%, and $\sim$30\% of the total strong lensing cross section respectively~\cite{42}. However, since we do not want to restrict our attention to specific cases and we expect the qualitative trends to remain the same, we use the SIS model and compare our results with those obtained by assuming a Navarro-Frenk-White (NFW) mass density profile~\cite{34} which is shallower than isothermal near the center and steeper in the outer regions,
\begin{equation}
\rho(x) =\frac{ \rho_{cr} \delta_{NFW}}{x(1+x)^2} \;\;\;,\;\;\; x = \frac{c}{r_{vir}}r =\frac{c}{\theta_{vir}}\theta 
\end{equation}
where $\rho_{cr}$ is the critical density at the epoch of the halo virialization. $\delta_{NFW}$ is related to $c$, the halo concentration parameter, by
\begin{equation}
\delta_{NFW} = \frac{200}{3}\frac{c^3}{\ln(1+c)-c/(1+c)}
\end{equation}
where $c$ can be calculated using the virial mass through a fit to simulations~\cite{43}
\begin{equation}
c(M,z) = \frac{9}{(1+z)}\left(\frac{M}{M'}\right)^{-0.13}
\end{equation}

\begin{figure}
\begin{tabular}{cc}
\hspace{-2cm}\includegraphics[width=280pt,height=250pt]{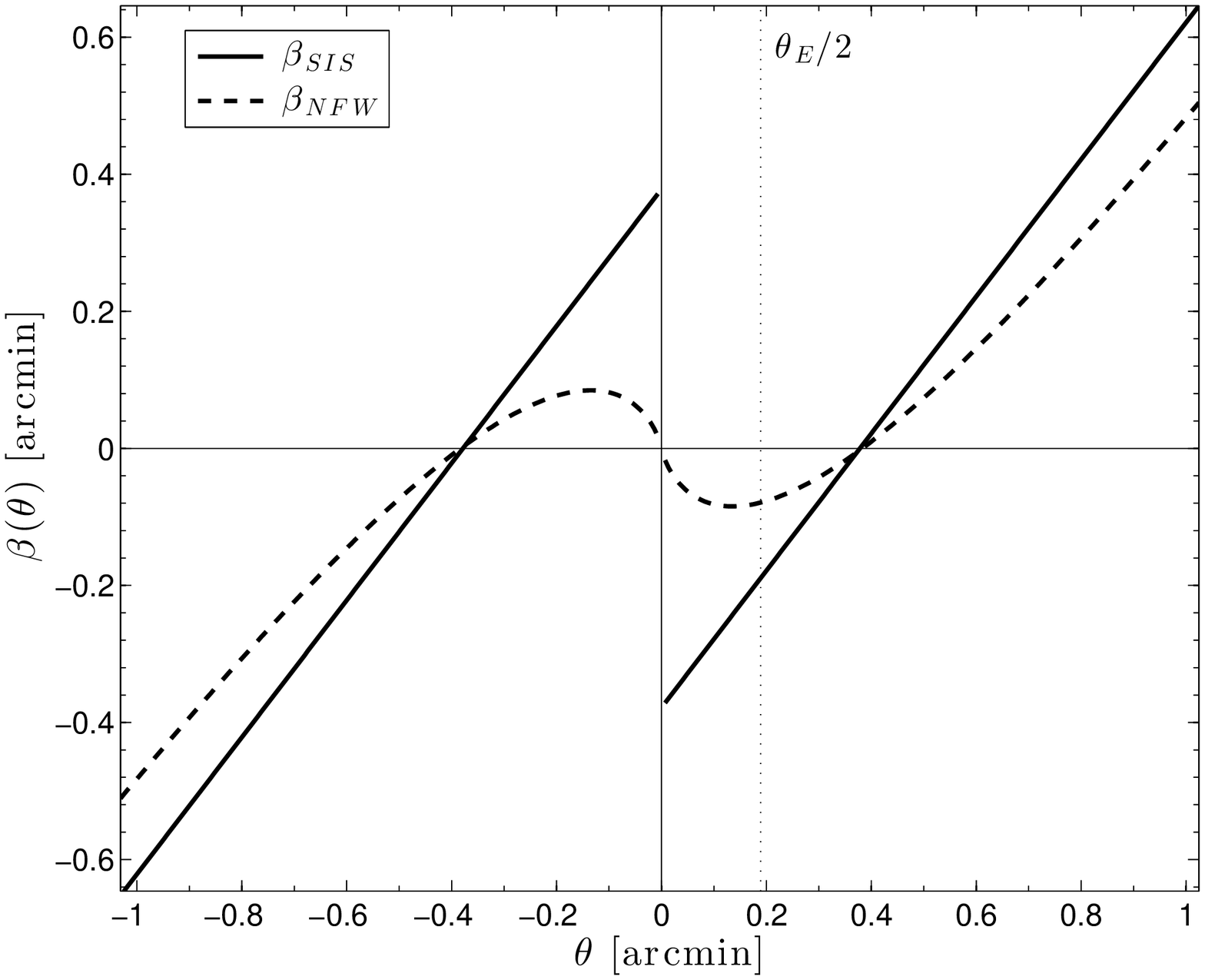}&
\hspace{-1cm}\includegraphics[width=280pt,height=250pt]{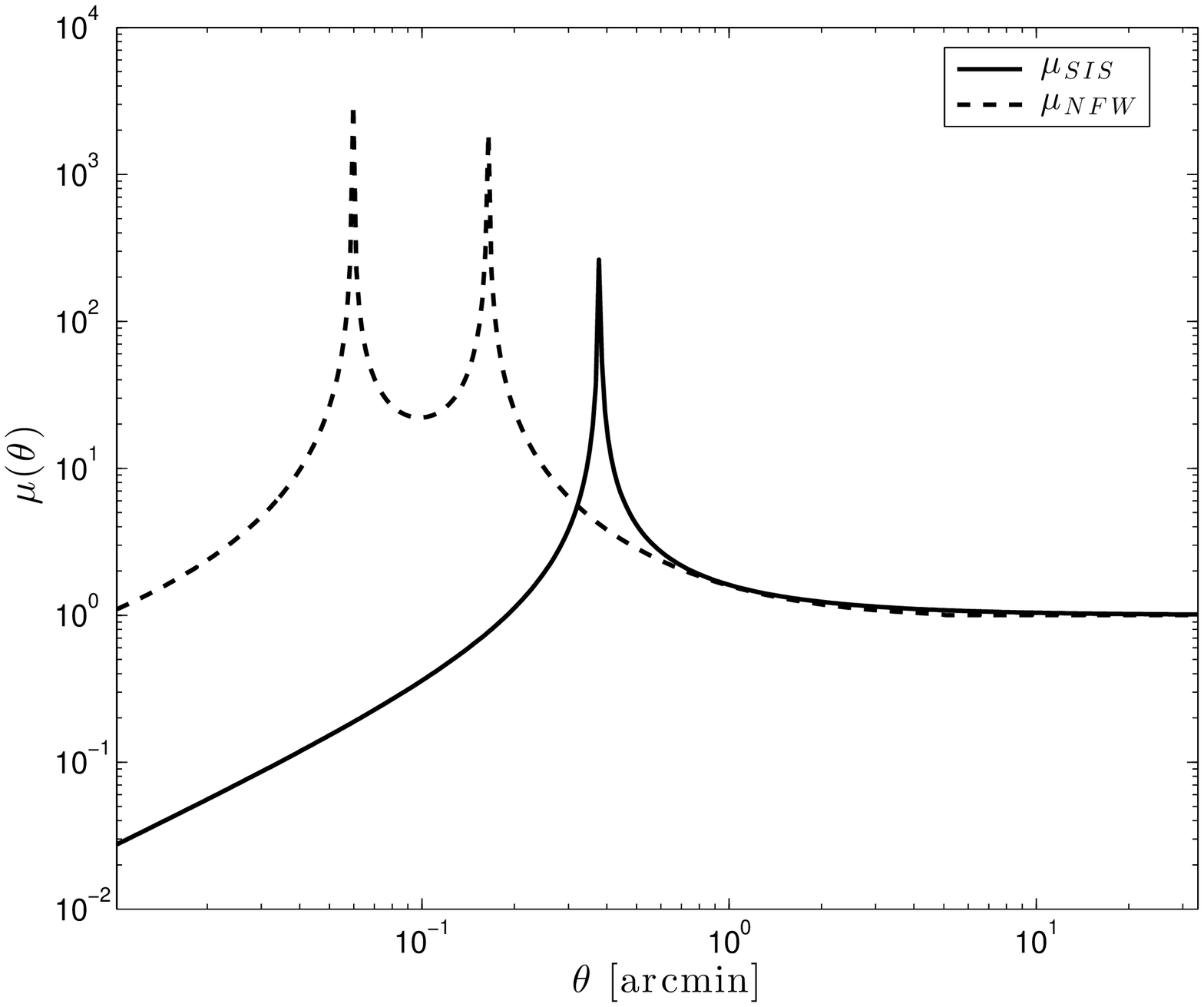}\\
\end{tabular}
\caption{\label{fig:i}\emph{Left:} Source position, $\beta$, as a function of image position, $\theta$ for SIS (solid) and NFW (dashed) lenses at redshift $z_L$ = 0.5 with halo mass $M$ = 10$^{14.9}$ M$_{\odot}$, ($\sigma_v$ = 1000 km s$^{-1}$) and a source redshift of $z_s$ = 8. A horizontal line of fixed source position $\beta$ may intersect each respective curve at multiple positions $\theta$, signaling the formation of multiple images. For the NFW lens, three images will form if $|\beta|$ $\leq$ $\beta_{cr}$, where $\beta_{cr}$ = -$\beta(\theta_{cr})$ and $\theta_{cr}$ $>$ 0 is determined by ($\left.d\beta/d\theta\right) |_{\theta = \theta_{cr}}$ = 0. For the SIS lens, two images will form if $|\beta| <$ $\theta_E$. \emph{Right:} Magnification as a function of angular separation $\theta$  from the cluster center assuming SIS (solid) and NFW (dashed) profiles with $z_L$ = 0.5 and $z_s$ = 8.}
\end{figure}

\noindent with $M'$ = 1.5$\times$10$^{13}h^{-1}M_{\odot}$. The virial radius of a halo at redshift $z$ depends on the halo mass as,
\begin{equation}
r_{vir} = 0.784\left(\frac{\Omega_m}{\Omega_m(z)}\frac{\Delta_c}{18\pi^2}\right)^{-1/3}\left(\frac{M}{10^8 M_{\odot}}\right)^{1/3}\left(\frac{1+z}{10}\right)^{-1}h^{-2/3} \text{ kpc} \;.
\end{equation}
In a universe with $\Omega_m + \Omega_\Lambda$ = 1, the virial overdensity at the collapse redshift has the fitting formula~\cite{44}
\begin{equation}
\Delta_c=18\pi^2+82d-39d^2
\end{equation}
with  $d$ = $\Omega_m(z)$+1 and 
\begin{equation}
\Omega_m(z) = \frac{\Omega_m(1+z)^3}{\Omega_m(1+z)^3+\Omega_\Lambda} \;.
\end{equation}

The lens equations for the NFW profile~\cite{45}, use the dimensionless surface mass density is
\begin{equation}
\kappa_{NFW}(x) = \frac{2\rho_{cr}\delta_{NFW}r_{vir}}{c \,\Sigma_{cr}}\frac{f(x)}{x^2-1}
\end{equation}
with
\begin{equation}
f(x) = \begin{cases}
1-\frac{2}{\sqrt{x^2-1}}\arctan{\sqrt{\frac{x-1}{x+1}}}\;, & \text{ x $>$ 1}\\
1-\frac{2}{\sqrt{1-x^2}}\arctan{\sqrt{\frac{1-x}{x+1}}}\;, & \text{ x $<$ 1}\\
0 \;,& \text{ x =1} \;.
\end{cases}
\end{equation}
The ray-tracing equation takes the form
\begin{equation}
\beta = \theta - \left(\frac{\theta_{vir}}{c}\right)^2\frac{m_{NFW}(c\theta/\theta_{vir})}{\theta}
\end{equation}
where the dimensionless mass in this case is
\begin{equation}
m_{NFW}(x) = \frac{4\rho_{cr}\delta_{NFW}r_{vir}}{c\,\Sigma_{cr}}g(x) \;\;\; \text{ with } \;\;\; g(x) = \ln{\frac{x}{2}}+1-f(x)
\end{equation}
and $\theta_{vir}$ = $r_{vir}$/$D_l$. As can be seen in Figure~\ref{fig:i}, an NFW profile lens will form three distinct images if $|\beta|$ $\leq$ $\beta_{cr}$, where $\beta_{cr}$ = -$\beta(\theta_{cr})$ and $\theta_{cr}$ $>$ 0 is determined by 
($\left.d\beta/d\theta\right) |_{\theta = \theta_{cr}}$ = 0; for all $|\beta|$ $\geq$ $\beta_{cr}$, a single image is formed.

The equations for $\kappa_{NFW}(x)$ and $m_{NFW}(x)$, used in conjunction with eq. \eqref{2.7}, yield the corresponding magnification factor for a NFW profile lens. The two spikes in $\mu_{NFW}(\theta)$ seen in Figure~\ref{fig:i} represent the tangential and radial critical curves where the magnification is formally infinite. The critical curves of the NFW lens are closer to the lens center than for the SIS lens and the image magnification decreases, approaching unity, more slowly away from the critical curves. 

To model the background galaxy population, we use the Schechter luminosity function,
\begin{equation}\label{2.22}
\phi(z,M)dM = 0.4\ln{10}\; \phi^*(z)\,10^{0.4\,(\alpha(z)+1)(M^*(z)-M)}e^{-10^{0.4(M^*(z)-M)}}
\end{equation}
where the parameters are the comoving number density of galaxies $\phi^*$, the characteristic absolute AB magnitude $M^*$, and the faint-end slope $\alpha$. (Note that we denote absolute AB magnitude in this paper as $M$.) The evolution of $\phi^*$ and $M^*$ as functions of redshift in the interval $z \geq$ 4 are taken as the central values of the fitting formulae provided in~\cite{46},
\begin{equation}
\phi^*(z) = (1.14 \pm 0.20)\times 10^{-3}10^{(0.003\pm0.055)(z-3.8)}\; \mathrm{Mpc^{-3}} \;,
\end{equation}
\begin{equation}
M^{*}(z) = (-21.02\pm 0.09) + (0.33\pm 0.06)(z-3.8) \;.
\end{equation}
~\cite{46} also provides a fitting formula describing the evolution of $\alpha$ as a function of redshift,
\begin{equation}\label{2.25}
\alpha(z) = (-1.73\pm0.05)+(-0.01\pm0.04)(z-3.8) \;. 
\end{equation}
Recent studies have investigated the form of the $z$ = 8 luminosity function by combining the faint-end results in~\cite{46} with improved constraints at the bright end~\cite{47,48,49,50}. There is very good agreement between the new results, with all studies converging on a steep faint-end slope of $\alpha\simeq$ -2.0. We therefore adopt the expression for $\alpha(z)$ in eq. \eqref{2.25} for all redshifts $z <$ 8  and use a faint-end slope of $\alpha$ = -2.02 for all higher redshifts, assuming that the slope remains unchanged for redshifts $z\geq$ 8~\cite{49}. This faint-end slope, along with the formulae describing the evolution of the LF, represent an extrapolation of the present LF results  ($z\sim$7-8) to even higher redshifts; the fall-off in UV luminosity at $z > $ 8 is still debated in the literature~\cite{21,49}. The results in this paper may therefore change as the evolution of the LF parameters $\phi^*$, $M^*$, and $\alpha$ as functions of redshift are modified in light of new observations. 
\\ \indent In the absence of a lensing object ($\mu$ = 1), the number of sources with redshift in the range $z_i < z < z_f$ seen in an angular region [$\theta_i$, $\theta_f$] about the optical axis is simply the number which falls in the angular region with an absolute magnitude less than the limiting absolute magnitude. This limiting magnitude is set either by the maximum intrinsic absolute magnitude associated with a star-forming halo, $M_{max}$, or, by what we denote as $M_{det}$, the absolute magnitude that a source at redshift $z$ must have to be above $df/d\nu_0$, the flux threshold set by the detector. This number is thus obtained by integrating the comoving number density $\phi(z,M)dM$  given by eq. \eqref{2.22} over the appropriate volume and magnitude range,
\begin{equation}
N^{\mathrm{\tiny{unlensed}}}(\theta_i, \theta_f, z_i,z_f) =2\pi \int_{\theta_i}^{\theta_f}d\theta^\prime\theta^\prime\;\;\frac{dN}{d\Omega}^{\mathrm{\tiny{unlensed}}}\!\!\!\!\!\! \!\!\!\!\!\! \!\!\! \!\!\!  (z_i,z_f)
\end{equation}
where
\begin{equation}
\frac{dN}{d\Omega}^{\mathrm{\tiny{unlensed}}}\!\!\!\!\!\! \!\!\!\!\!\! \!\!\! \!\!\!  (z_i,z_f) = \frac{c}{H_0} \int_{z_i}^{z_f} \frac{dz}{E(z)} D^2_{s,com}(z)\int_{-\infty}^{\mathrm{Min}[M_{max},M_{det}(z)]} \!\!\! \!\!\! \!\!\! \!\!\! \!\!\!\!\!\!  \!\!\! \!\!\!  \!\!\!  \!\!\!\!\!\!      dM\: \phi(z,M) \;.
\end{equation}
and $D_{com}$ is the comoving angular diameter distance,  $D_{com}(z)$ = $D_{A}(1+z)$. In the presence of a lens, the magnification due to gravitational lensing has two effects on background point sources: their surface number density is diluted by a factor of $\mu$,
\begin{equation}
n^{\mathrm{\tiny{obs}}}(\theta) = n/\mu(z,\theta) \;,
\end{equation}
 and their luminosities are simultaneously  magnified by the same factor,
\begin{equation}
L^{\mathrm{\tiny{obs}}}(\theta) = \mu(z,\theta) L \;\;\;\rightarrow\;\;\; M^ {\mathrm{\tiny{obs}}} = M + 2.5\log{\mu(z,\theta)} \;.
\end{equation}
Furthermore, in the case of strong lensing, multiple images will often be produced by SIS and NFW profile lenses depending on the source position, $\vec{\beta}$. When $|\beta|< \theta_E$, an SIS lens will form two images with a splitting angle of $\Delta\theta_{SIS}$ = 2$\theta_E$. Similarly, an NFW profile lens will form three distinct images if $|\beta| \leq \beta_{cr}$; however, only two of those images will lie in the region $|\theta| \geq \theta_E/2$ (Figure~\ref{fig:i}), the region of interest in the following section. In general, the splitting angle $\Delta\theta$ between these two outside images is insensitive to the value of $\beta$ and is approximately given by~\cite{51}
\begin{equation}
\Delta\theta_{NFW} \approx \Delta\theta(\beta=0) = 2\theta_0, \text{for  $|\beta| < \beta_{cr}$}
\end{equation}
where $\theta_0$ is the positive root of $\beta(\theta)$ = 0. Consequently, the total number of sources detected in a lensed field takes the following modified form,
\begin{eqnarray}
N^{\mathrm{\tiny{lensed}}}(\theta_f, z_i,z_f) &=& 2\pi \int_{\theta_E/2}^{\Delta\theta/2}d\theta^\prime\theta^\prime\; \frac{dN}{d\Omega}^{\mathrm{\tiny{lensed}}}\!\!\!\!\!\!\!\!\!\!\!\!(\theta^\prime,z_i,z_f) \nonumber\\
&+& 2\pi \int_{\Delta\theta/2}^{\Delta\theta-\theta_E/2}d\theta^\prime\theta^\prime\;\; \mathrm{max}\left[\;0 \;,\; \frac{dN}{d\Omega}^{\mathrm{\tiny{lensed}}}\!\!\!\!\!\!\!\!\!\!\!\!(\theta^\prime,z_i,z_f) -  \frac{dN}{d\Omega}^{\mathrm{\tiny{lensed}}}\!\!\!\!\!\!\!\!\!\!\!\!(|\theta^\prime - \Delta\theta|,z_i,z_f)\right]\nonumber\\
&+&2\pi \int_{\Delta\theta-\theta_E/2}^{\theta_f}d\theta^\prime\theta^\prime\; \frac{dN}{d\Omega}^{\mathrm{\tiny{lensed}}}\!\!\!\!\!\!\!\!\!\!\!\!(\theta^\prime,z_i,z_f) 
\end{eqnarray}
where
\begin{equation}
\frac{dN}{d\Omega}^{\mathrm{\tiny{lensed}}}\!\!\!\!\!\!  (\theta^\prime,z_i,z_f) = \frac{c}{H_0} \int_{z_i}^{z_f} \frac{dz}{E(z)} D^2_{s,com}(z)\frac{1}{\mu^2(z,\theta^\prime)} \int_{-\infty}^{\mathrm{min}[M_{max},M_{det}(z)+2.5\log{\mu(z,\theta^\prime)}]} \!\!\! \!\!\!  \!\!\! \!\!\! \!\!\!  \!\!\!  \!\!\! \!\!\!  \!\!\!  \!\!\! \!\!\!\!\!\!\!    \!\!\!\!\!\!\!\!\!\!\!  \!\!\!\!\!\!\!\!\!\!\!dM\: \phi(z,M) \;.
\end{equation}

\begin{figure}[H]\vspace{-1cm}
\centering
\begin{tabular}{cc}
\vspace{-.5cm}\hspace{-1cm}\includegraphics[width=460pt,height=285pt]{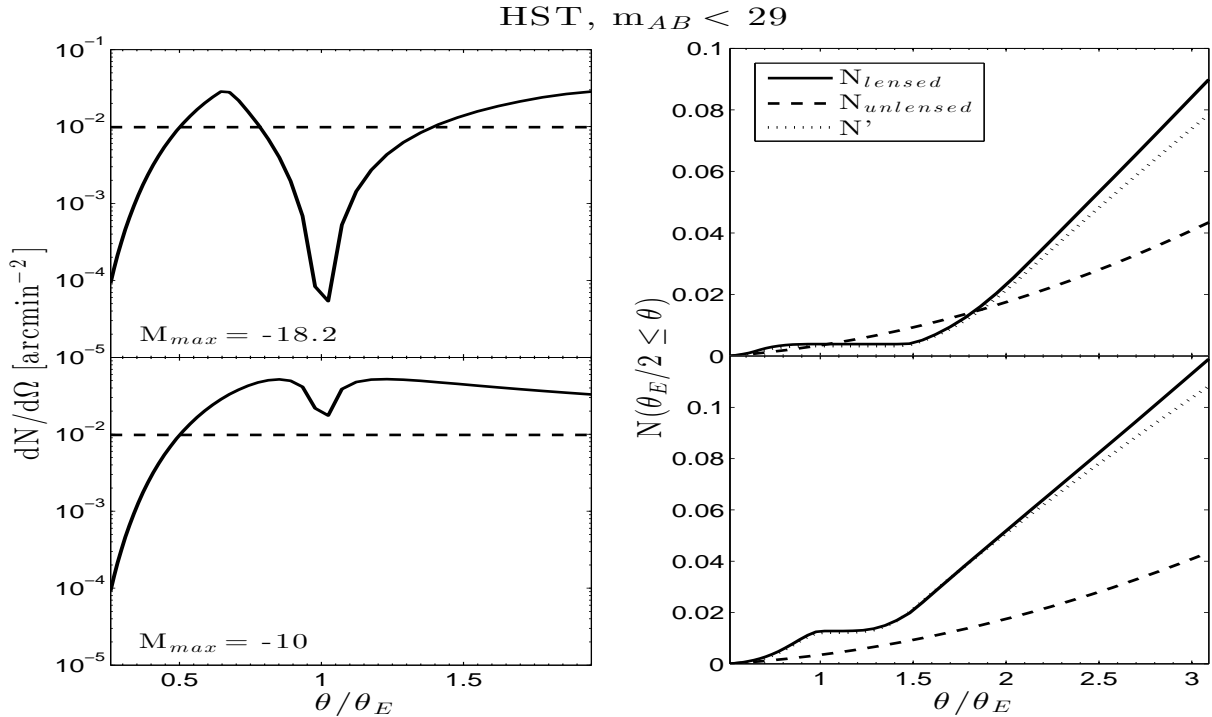}\\
\caption{\label{fig:ii} Predictions for observations by HST ($df/d\nu_0 \sim$ 9 nJy) of sources at redshifts z $\gtrsim$ 13 behind a SIS cluster lens with $\sigma_v \approx$ 1000 km s$^{-1}$ at $z_L$ = 0.5. \emph{Left Panel}: Radial dependence of the number of sources detected in concentric annular cells \emph{Right panel:} Cumulative number of detected sources, starting from an angular radius $\theta_E$/2 and extending outward to an angular radius $\theta$.  The solid and dashed lines correspond to the lensed and unlensed numbers respectively. The dotted line in the right panel corresponds to $N'$, the cumulative number of magnified sources that were lifted over the detection threshold of the given instrument; this represents the component of $N_{lensed}$ that would remain undetected without the aid of lensing. The top panel assumes a maximum intrinsic absolute AB magnitude of -18.2 mag while the bottom panel assumes $M_{max}$ = -10 mag.}
\end{tabular}
\end{figure}

\section{Results}
\label{3}
We now apply the general relations discussed above to lensing of background sources by a lens positioned at a redshift $z_L$ = 0.5. This redshift is chosen to be consistent with the average redshift of the galaxy clusters centered in the six deep fields of the HST Frontier Fields program. In all the calculations presented below, we take $z_f$ = 16 as the upper bound on the redshift range. Beyond this redshift, $z_f >$ 16, the results remain the same (within a precision of one part in ten-thousand); the numerical results drop by $\lesssim$ 4\% if instead we assume the first galaxies formed at $z_f$ = 13. Although we focus on the lensing effect due to galaxy clusters ($\sigma_v \approx$ 1000 km/s), we also include the results obtained when considering lensing by galaxy groups ($\sigma_v \approx$ 500 km/s).  We restrict our attention to the flux thresholds set by the WFC3 aboard the HST and  the NIRCam imager of JWST. The Frontier Fields program achieves AB $\approx$ 28.7-29 mag optical (ACS) and NIR (WFC3) imaging, corresponding to flux limits of $df/d\nu_0 \sim$ 9-12 nJy for a 5$\sigma$ detection of a point source after $\sim$ 10$^5$ s.  Medium-deep ($m_{lim} \simeq$ 29.4 mag) and ultra-deep ($m_{lim} \simeq$ 31.4 mag) JWST surveys, (corresponding

\begin{figure}[H]\vspace{-1cm}
\centering
\begin{tabular}{cc}
\vspace{-0.4cm}\includegraphics[width=460pt,height=285pt]{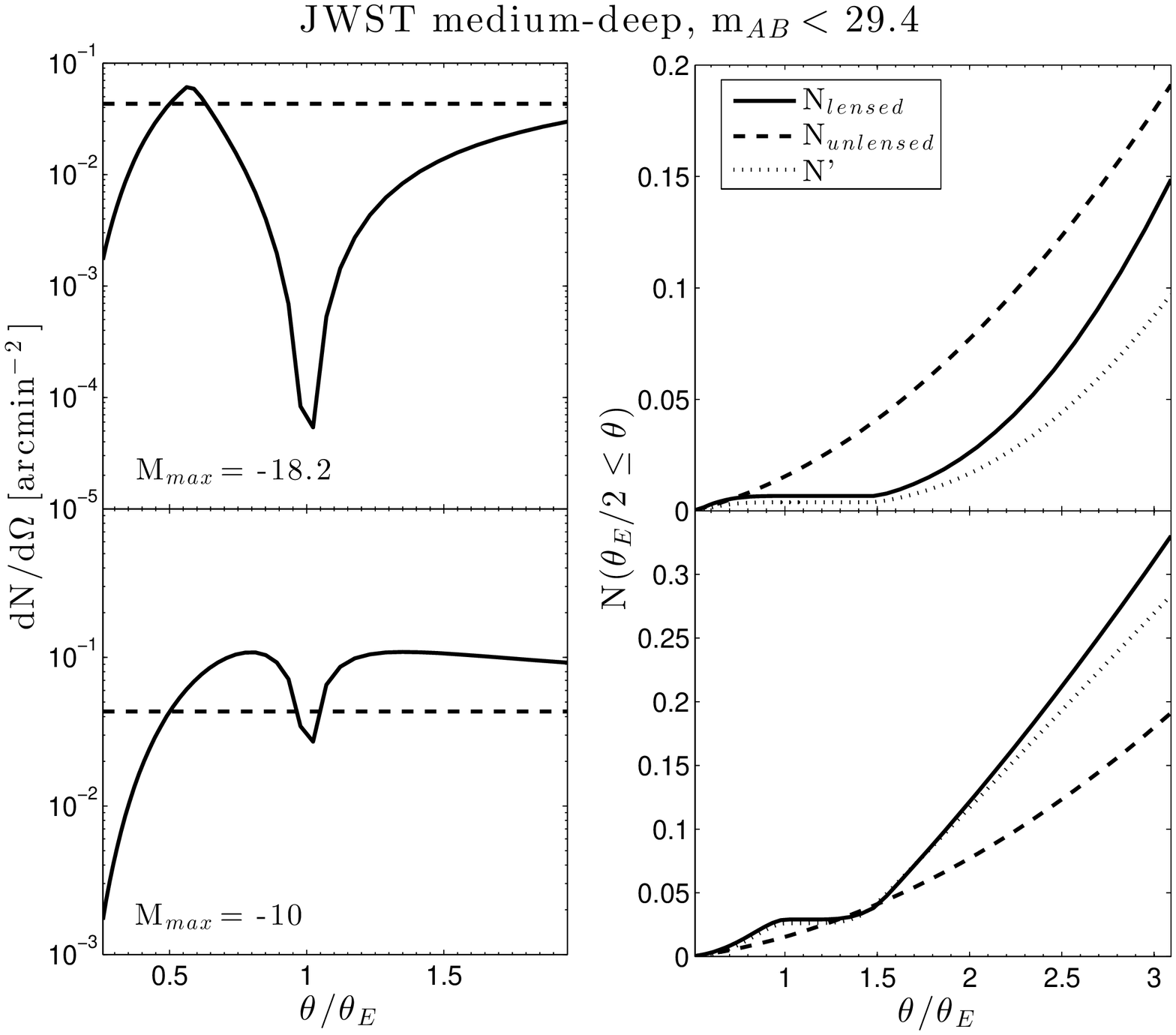}\\
\vspace{-0.8cm}\includegraphics[width=460pt,height=285pt]{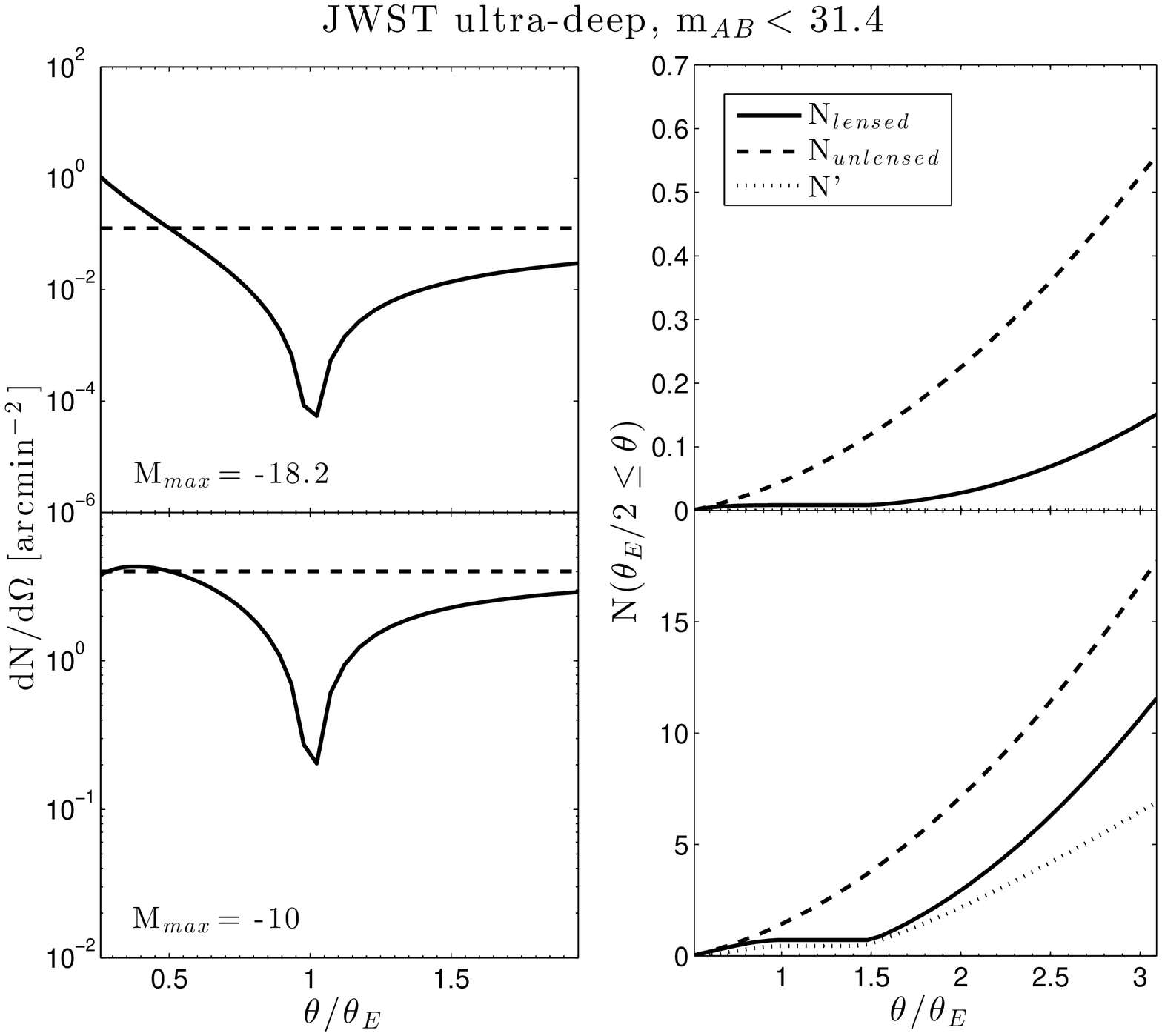}\\
\caption{\label{fig:iii} Predictions for observations in JWST's medium-deep ($df/d\nu_0 \sim$ 6 nJy) and ultra-deep  ($df/d\nu_0 \sim$ 1 nJy) fields of sources at redshifts z $\gtrsim$ 13 behind a SIS cluster lens with $\sigma_v \approx$ 1000 km s$^{-1}$ at $z_L$ = 0.5. \emph{Left Panel}: Radial dependence of the number of sources detected in concentric annular cells. \emph{Right panel:} Cumulative number of detected sources, starting from an angular radius $\theta_E$/2 and extending outward to an angular radius $\theta$.  The solid and dashed lines correspond to the lensed and unlensed numbers respectively. The dotted line in the right panel corresponds to $N'$, the cumulative number of magnified sources that were lifted over the detection threshold of the given instrument due to lensing by an SIS profile lens; this represents the component of $N_{lensed}$ that would remain undetected without the aid of lensing. For each deep-field, the top panel assumes a maximum intrinsic absolute AB magnitude of -18.2 mag while the bottom panel assumes $M_{max}$ = -10 mag.}
\end{tabular}
\end{figure}

\noindent to a 5$\sigma$ detection of a point source after 10$^4$ and 3.6$\times$10$^5$ s of exposure respectively), will detect fluxes as low as $\sim$ 6 and 1 nJy, respectively. 

The radial dependence of the number of background sources with redshifts $z \geq$ 13  detected by HST and JWST behind a SIS lens at $z_L$ = 0.5 is depicted in the left panels of Figures~\ref{fig:ii} and~\ref{fig:iii}, respectively. At large distances from the lens center ($\theta/\theta_E \gg$ 1), the magnification factor approaches unity  and the number of detected sources per annular ring converges to the constant number that would be observed in the absence of a lens (dotted line). For an image at $|\theta| < \theta_E$/2, $|\mu|$ is smaller than unity, and the source luminosity is demagnified relative to the unlensed luminosity while the surface number density is amplified by the same factor. As $\theta$ approaches the Einstein angle, the magnification diverges, allowing small sources perfectly aligned with the lens center to form an "Einstein ring" and otherwise, causing the number density of observed sources to plummet. (This phenomeneon corresponds to the sharp drop in $dN/d\Omega$ at $\theta/\theta_E$ = 1). At image distances larger than the Einstein angle, $\mu$ converges back to unity, resulting in magnified luminosities and diluted number densities that gradually reduce to their unlensed values. The overall magnification bias depends on which of the two magnification effects wins out: under circumstances where the number of  magnified sources lifted over the detection threshold outweighs the simultaneous dilution of the number density of sources in the sky, there is a positive magnification bias and $\left(dN/d\Omega\right)_{lensed} > \left(dN/d\Omega\right)_{unlensed}$. Conversely, when the reduction in number density dominates over luminosity magnifications, a negative magnification bias results and $\left(dN/d\Omega\right)_{lensed} < \left(dN/d\Omega\right)_{unlensed}$ in those regions.

The plots of $dN/d\Omega$ and $N(\theta_E/2<\theta)$ (the cumulative number of sources observed starting from an angular radius of $\theta_E/2$ and extending outward to an angular radius $\theta$) in Figures~\ref{fig:ii} and~\ref{fig:iii} highlight the sensitivity of the magnification bias to the different model parameters. The expected magnification bias effect on the source counts observed around a lensing cluster depends strongly on the flux threshold of the instrument used for the survey and its strength relative to the characteristic magnitude of the galaxy sample. If the instrumental detection threshold places the observer  in the exponential drop-off region of the luminosity function of a galaxy sample, ($M_{det}$(z,f$_{min}$) $<$ $M^*(z)$), lensing will significantly increase the number of observed sources when it pushes the detection threshold to fainter values of $M$, resulting in a positive magnification bias. This is the case with detections of $z \geq$ 13 galaxies in HST and medium-deep JWST surveys. On the other hand, in an ultra-deep

\noindent JWST survey, where the instrumental limiting magnitude is fainter than the characteristic magnitude, (placing the observer in the power-law region of the Schechter function), pushing the threshold to fainter magnitudes does not result in the inclusion of a substantial population of otherwise undetectable sources; the diluting effect of lensing therefore wins out and a deficit in the total source count is consistently observed (bottom panel, Figure ~\ref{fig:iii}). Source counts of lower redshift galaxy populations in HUDF and medium-deep JWST surveys suffer from this negative magnification bias as well. 

However, note that even in these instances where gravitational lensing reduces the \emph{total} source count, a significant fraction of the sources that are observed in the lensed field belong to a population of galaxies that, without lensing, lie below the survey limit (dotted lines in right panels of  Figures~\ref{fig:ii} -~\ref{fig:iii}). By lifting these galaxies over the instrumental detection threshold, lensing may help constrain another model parameter, $M_{max}$, the maximum intrinsic absolute magnitude of the background galaxy population.

\begin{figure}[H]\vspace{-1cm}
\centering
\begin{tabular}{cc}
\vspace{-.3cm}\hspace{-2cm}\includegraphics[width=550pt,height=480pt]{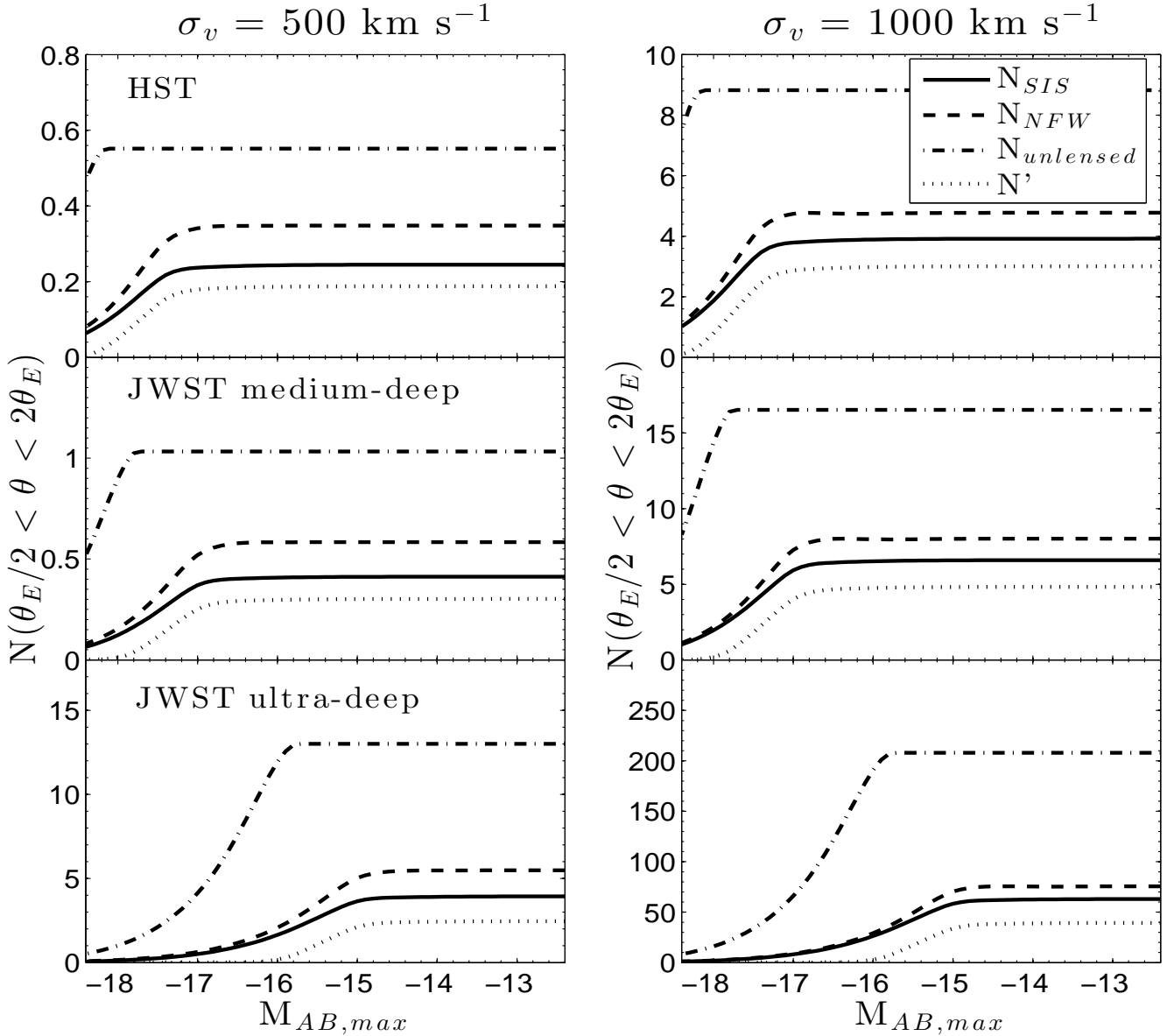}\\
\end{tabular}
\caption{ \label{fig:iv}Expected number of sources with redshift $\bm{z \gtrsim 8}$ integrated over the region $\theta_E/2 \leq \theta \leq 2\theta_E$ as a function of the maximum intrinsic absolute magnitude, $M_{AB,max}$, of the background sources.The solid and dashed lines respectively show the expected numbers due to lensing by a SIS and NFW-profile lens respectively. The dash-dot line represents the expected numbers in the unlensed case. The dotted line represents the number of sources in the lensed field that have been lifted over the given instrumental detection threshold due to lensing by an SIS profile lens. The top, center, and bottom panels show the expected results given a flux threshold of 9 (HST), 6 (JWST medium-deep), and 1 nJy (JWST ultra-deep) in the case where galaxy groups (left panel) and clusters (right panel) are used as lenses with corresponding Einstein angles 0.095', and 0.38'.}
\end{figure}

\begin{figure}[H]\vspace{-1cm}
\centering
\begin{tabular}{cc}
\hspace{-2cm}\vspace{-.5cm}\includegraphics[width=550pt,height=480pt]{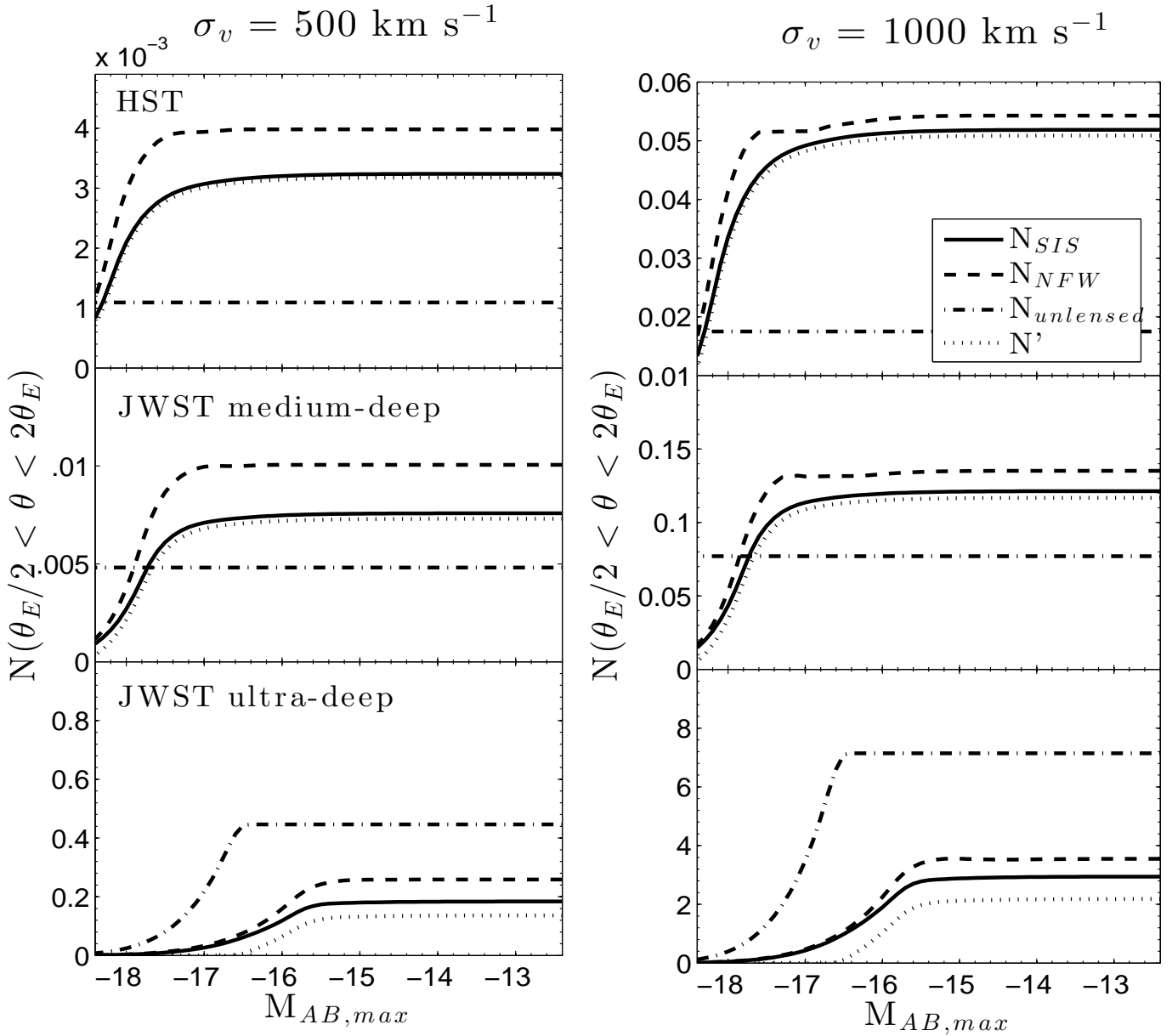}\\
\end{tabular}
\caption{ \label{fig:v}Expected number of sources with redshift $\bm{z \gtrsim 13}$ integrated over the region $\theta_E/2 \leq \theta \leq 2\theta_E$ as a function of the maximum intrinsic absolute magnitude, $M_{AB,max}$, of the background sources.The solid and dashed lines respectively show the expected numbers due to lensing by a SIS and NFW-profile lens, respectively. The dash-dot line represents the expected numbers in the unlensed case. The dotted line represents the number of sources in the lensed field that have been lifted over the given instrumental detection threshold due to lensing by an SIS profile lens. The top, center, and bottom panels show the expected results given a flux threshold of 9 nJy (HST), 6 nJy (JWST medium-deep), and 1 nJy (JWST ultra-deep) in the case where galaxy groups (left panel) and clusters (right panel) are used as lenses with corresponding Einstein angles 0.095', and 0.38'.}
\end{figure}

Figures ~\ref{fig:iv} -~\ref{fig:v} depict the expected source count integrated over the range of angular distances $\theta$ $\epsilon$ [$\theta_E$/2, 2$\theta_E$] from the lens center, as a function of the $M_{max}$, assuming a SIS

\noindent (solid) and NFW (dashed) profile.  Images in this region fall far enough away from the lens that their detectability is not compromised by the brightness of the foreground lens, yet close enough that the magnification bias introduced by $\mu(\theta)$ has a noticable effect on the

\noindent source count. These plots demonstrate that the degree to which the number count varies as a function of $M_{max}$ at the fainter end of the considered range relies significantly on the presence of a magnification bias. In the case of a blank field, the number count of sources detected by an instrument grows insensitive to $M_{max}$ if this faint-magnitude cut-off is fainter than $M_{det}$; since sources with magnitudes fainter than $M_{det}$ cannot be observed without the aid of gravitational lensing, the unlensed number count, $N_{unlensed}$, plateaus for values of $M_{max} \gtrsim M_{det}$ and the presence of these faint galaxies remains unverifiable. Therefore, an instrument with a flux threshold of $\sim$9 nJy, such as HST, cannot confirm the existence of sources at redshifts $z \gtrsim$ 8 and $z \gtrsim$ 13 with absolute magnitudes fainter than -18.1 and -18.9 mag, respectively. Similarily, the medium-deep JWST survey loses sensitivity to sources fainter than -17.7 ($z\geq$ 8) and -18.5 mag ($z\geq$ 13) while the ultra-deep survey will not detect sources fainter than -15.7 ($z\geq$ 8) and -16.5 mag ($z\geq$ 13) when observations are made in a blank field. This sensitivity to the intrinsic faint-magnitude cut-off significantly improves when considering $N_{lensed}$, the number of high-redshift sources one expects to observe behind a lensing group or cluster. Although the gain in depth  does not balance the dilution of sources in most instances, particularly when observing sources at redshifts $z < $ 13 with HST and JWST, it permits the detection of sources fainter than $M_{det}$ and thus allows the lensed number count to remain sensitive to $M_{max}$, down to values much fainter than was the case for the unlensed number count. In particular, the number count of $z \gtrsim$ 8 galaxies lensed by a foreground cluster modeled as a SIS lens, can be used to infer the intrinsic faint-magnitude cut-off of the Schecter function up to values as faint as $M_{max} \sim$ -17.8, -17.3, and -15.0 mag ($L_{min} \sim$ 5.8$\times$10$^{27}$, 3.6$\times$10$^{27}$, and 4.4$\times$10$^{26}$ erg s$^{-1}$Hz$^{-1}$) in the HUDF, medium-deep, and ultra-deep JWST surveys, respectively (Figure~\ref{fig:iv}, right panel). Similarly, observations of $z \gtrsim$ 13 galaxies in ultra-deep JWST surveys can yield an estimate of $M_{max}$ for values as faint as -16.1 mag ($L_{min} \sim$ 1.2$\times$10$^{27}$ (Figure~\ref{fig:v}, right panel). Modeling the cluster as a NFW lens instead results in the same constraints on $M_{max}$ and changes the magnitude of $N_{lensed}$ by at most $\sim$ 20\% compared to the numbers obtained for the SIS lens.

\begin{figure}[H]\vspace{-1cm}
\centering
\begin{tabular}{cc}
\includegraphics[width=350pt,height=300pt]{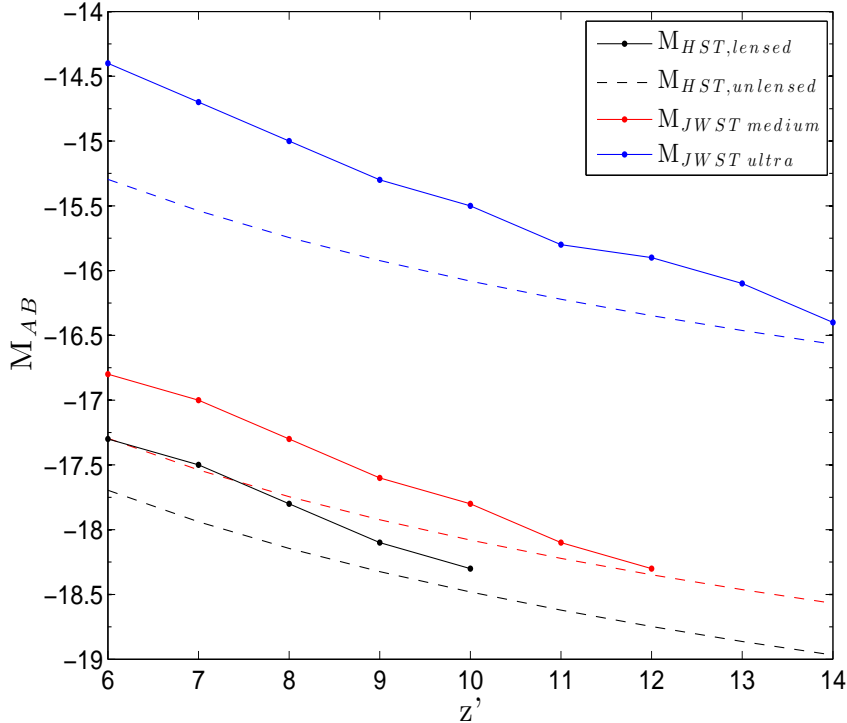}\\
\caption{\label{fig:vi} The lower bound on $M_{max}$ obtained with and without the lensing effect. The dashed lines represent $M_{det}(z', f_{min})$, the absolute AB magnitude of the faintest galaxy at redshift $z$ $\geq$ $z'$ that can be detected in the HUDF (black), JWST medium-deep (red), and JWST ultra-deep (blue) surveys. Without the aid of lensing, these detection thresholds set the lower bound on the intrinsic maximum absolute magnitude of the Schecter luminosity function. The solid lines represent the tightest constraint on $M_{max}$  that can be obtained using the lensed number count of galaxies in each of the respective surveys (solid lines in Figures~\ref{fig:iv} and~\ref{fig:v}).}
\end{tabular}
\end{figure}

Given that the number of galaxies in a field of observation follows a Poisson distribution and that the sample size is large (which is expected in fields centered on strong lensing galaxy clusters), the 1-$\sigma$ confidence interval around the measured count translates into a range of inferred $L_{min}$ values. Using the appropriate plots in Figures~\ref{fig:iv} and~\ref{fig:v}, one can therefore identify the most likely value of $M_{max}$ to within the Poisson uncertainty of the observed number count, $\sqrt{N_{obs}}$. An observed number count with an error bar that lies outside the model, (i.e. $N_{obs}$ + $\sqrt{N_{obs}}$ exceeds the maximum number count accommodated by the model), will subsequently yield only a lower bound on the intrinsic faint-magnitude cut-off. Figure~\ref{fig:vi} compares the constraint on $M_{max}$ that may be obtained by observing galaxy populations within a given range of redshifts with HST and JWST in blank versus lensed fields. When observing $z\geq$ 6 galaxies, the intrinsic maximum magnitude can be inferred up to values as faint as -14.4 mag (2.5$\times$10$^{26}$ erg s$^{-1}$Hz$^{-1}$) with JWST.

In addition to constraining the value of $L_{min}$, Figures~\ref{fig:iv} and~\ref{fig:v} also demonstrate the instances in which gravitational lensing improves the detection of high redshift sources as well as the magnitude of the bias in each case.  Gravitational lensing by a group or cluster modeled as an SIS lens reduces the number of detected sources with redshift $z \gtrsim$ 8 by a factor of $\sim$ 2-3 in HST and JWST observations. However, even though the total source count is reduced in these cases, the majority of the sources that are observed in the lensed field are galaxies that, without lensing, would remain undetectable. In particular, nearly  $\sim$ 73-76\% of the $z \geq$ 8 galaxies observed in HUDF and JWST surveys have been lifted over the respective instrumental detection thresholds (dotted lines). At higher redshifts, the magnification bias transitions from negative to positive and enhances the number of  $z\gtrsim$ 13 sources detected behind a lensing group or cluster by a factor of $\sim$ 3 and 1.5 in HUDF and JWST medium-

\noindent field surveys, respectively. Of these observed sources, nearly $\sim$ 96-98\% of them lie below the survey limit and would thus remain undetected without the aid of gravitational lensing

\section{Discussion}
\label{4}
In this paper, we studied the effects of gravitational lensing on the source count of high redshift objects as observed by both the HST Frontier Fields and JWST. Although lensing magnifies the background sources, effectively lowering the flux threshold above which they can be detected, it simultaneously dilutes the apparent number density of sources on the sky. We found that the details of whether the number counts of distant background sources seen through a foreground gravitational lens are enhanced or reduced depends on several parameters characterizing the system. Using the axially symmetric SIS and NFW mass density profiles to model a lens residing at a redshift $z_L$ = 0.5, we explored how the magnification bias varied with the velocity dispersion of the lens ($\sigma_v$), the angular distance from the lens ($\theta$), the photometric sensitivity of the instrument ($df/d\nu_0$), the redshift of the background source population, and the intrinsic faint-magnitude cut-off characterizing the population ($M_{max}$). We found that when observing sources at redshifts $z \gtrsim$ 8, lensing by a group or cluster will reduce the number of detected sources by a factor of $\sim$ 2-3 in both HST and JWST observations. The magnification bias transitions from negative to positive only when considering higher redshift galaxies; in particular, the bias will enhance the number of sources at redshifts $z \gtrsim$ 13  behind a lensing group or cluster by a factor of $\sim$ 3 and 1.5 in HUDF and JWST medium-deep surveys, respectively.

Although the gain in depth  does not balance the simultaneous dilution of sources in most instances, it permits the detection of sources fainter than $M_{det}$ and thus allows the lensed number count to remain sensitive to $M_{max}$, down to values much fainter than was the case for the unlensed number count. In particular, the number count of $z \gtrsim$ 8 galaxies lensed by a foreground cluster, can be used to infer the intrinsic maximum magnitude of the Schecter function up to values as faint as $M_{max} \sim$ -17.8 and -15 mag ($L_{min} \sim$ 5.8$\times$10$^{27}$ and 4.4$\times$10$^{26}$ erg s$^{-1}$Hz$^{-1}$) in the HUDF and ultra-deep JWST surveys, respectively, within the range bracketed by existing theoretical models~\cite{31,33}. Similarly, observations of $z \gtrsim$ 13 galaxies in ultra-deep JWST surveys can yield an estimate of $M_{max}$ for values as faint as -16.1 mag ($L_{min} \sim$ 1.2$\times$10$^{27}$ erg s$^{-1}$Hz$^{-1}$).

\section{Acknowledgements}
\label{5}
We thank Dan Stark for helpful comments on the manuscript. This work was
supported in part by NSF grant AST-0907890 and NASA grants NNX08AL43G and
NNA09DB30A.This material is based upon work supported by the National Science Foundation Graduate Research Fellowship under Grant No. DGE1144152. Any opinion, findings, and conclusions or recommendations expressed in this material are those of the authors(s) and do not necessarily reflect the views of the National Science Foundation.

\end{document}